\documentstyle[aps,12pt,epsfig]{revtex}  
\begin{document}
\draft
\title{Spectator and participant decay in
 heavy ion collisions} 
\author{T.~Gaitanos, H.~H.~Wolter}
\address{Sektion Physik, Universit\"{a}t M\"{u}nchen, 
         Am Coulombwall 1, D-85748 Garching, Germany}
\author{C.~Fuchs}
\address{Institut f\"ur Theoretische Physik, Universit\"at T\"ubingen 
    D-72076 T\"ubingen, Germany }
\maketitle
\begin{abstract}
We analyze the thermodynamical state of nuclear matter in transport 
calculations of heavy--ion reactions. In particular we determine 
temperatures and radial flow parameters 
from an analysis of fragment energy spectra  and 
compare to local microscopic temperatures obtained from 
an analysis of local momentum space distributions. 
The analysis shows that the spectator reaches an equilibrated freeze-out 
configuration which undergoes simultaneous fragmentation. The fragments 
from the participant region, on the other hand, do {\it not} seem to 
come from a common fragmenting source 
in thermodaynamical equilibrium. \\ \\
PACS number(s): 25.75.-q,25.70.Mn
\end{abstract}

One of the major interests in the study of intermediate energy heavy--ion 
collisions is the understanding of the multifragmentation phenomenon and its 
connection with liquid--gas phase transitions \cite{poch95}. For this it has to 
be assumed that in a heavy ion collision at some 
stage a part of the system is both in thermodynamical equilibrium and instable. 
Such a configuration is often termed a freeze-out configuration. The 
multifragmentation process would reflect the parameters of this source, i.e. its 
temperature, density, and perhaps collective (radial) flow pattern. 
Experimentally the fragment energy spectra are described in terms of such 
freeze-out models to extract these parameters. On the other hand, information on 
these quantities is also sought from other observables, in 
particular from isotope ratios of fragments, and from excited state populations. 
In the past the conclusions drawn from these different sources have often been 
in conflict with each other.

One way to study whether this szenario is applicable is to analyze the results 
of transport calculations of heavy ion collisions 
\cite{aichelin,fuchs95,nemeth97,hombach98,daffin96}. 
Since such calculations reproduce 
reasonably well the asymptotic observables it should be meanigful to look also 
into their intermediate time behaviour to see whether, where, and when such 
freeze-out configurations exist. In ref. \cite{fuchs97} a method was developed 
to determine 
thermodynamical variables {\it locally} by analyzing the local momentum 
distribution even in the presence of possible anisotropies.  

In the present work we apply this analysis to intermediate energy collisions of 
$Au + Au$, which were also studied extensively experimentally  
\cite{eos95,reisdorf,reisdorf2}. We want to establish whether the concept 
of a freeze-out configuration 
used in the experimental analysis of  fragment kinetic energy spectra is 
supported in 
realistic transport calculations. Fragments are described using a coalescence 
algorithm as detailed later. 
We study, in particular,  the participant and 
spectator regions which 
represent rather clean thermodynamical situations in a heavy ion collision. In 
this way we also try to clarify some of the discrepancies between different 
methods of temperature definition.

We base our investigation on relativistic transport calculations of the 
Boltzmann-Nordheim-Vlasov type. In this work we use, in 
particular, the relativistic Landau-Vlasov approach which was described in 
detail in ref.\cite{fuchs95}. It uses Gaussian test particles in coordinate and 
momentum space and thus 
allows to construct locally a smooth momentum distribution. For the self 
energies in the transport calculation we have adopted the non-linear 
parametrization NL2 \cite{NL2}. In ref. \cite{gait99} we compared this 
parametrization to more realistic non-equilibrium self energies based on 
Dirac-Brueckner calculations. With respect to the thermodynamical variables 
discussed here, we found no essential differences between the two models, 
and thus we use the simpler NL2 here. A similar analysis with respect to fragment 
kinetic energy spectra in central collisions has been performed previously 
in ref. \cite{daffin96} in the framework of non-relativistic transport theory with 
special emphasis on the dependence on assumptions of the equation of state and in-medium 
cross sections. Here also a weak dependence of radial flow observables 
on particular choices of the microscopic input was found. 

The microscopic determination of a local temperature 
from the phase space distribution was discussed in 
detail in refs. \cite{fuchs97}. Thus we only briefly review the procedure here. 
The local momentum distribution obtained
from a transport calculation is subjected to a fit in terms of covariant 
hot Fermi--Dirac distributions of the form
\begin{equation}
n(x,{\vec k},T) = 
\frac{1}{ 1+ \exp\left[ - (\mu^* - k_{\mu}^* u^\mu )/T \right]}
\label{fermi}
\end{equation}
with the temperature $T$, the effective chemical potential $\mu^* (T)$ 
and $ k_{0}^* = E^* = \sqrt{ {\vec k}^{*2} + m^{*2} }$. 
For vanishing temperature eq.(\ref{fermi}) includes the limit of a sharp 
Fermi ellipsoid with $\mu^* = E_{\rm F} = \sqrt{k_{{\rm F}}^2 + m^{*2}}$. 
The local streaming 
four-velocity $u_\mu$ is determined from the local 4-current $j^{\mu}$ as 
$u^{\mu} = j^{\mu}/\rho_0$, where $\rho_0=\sqrt{j_\mu j^\mu}$ is the 
local invariant density. Then 
the temperature $T$ is the only fit parameter to be directly 
determined from the phase space distribution. In this procedure the 
effect of the potential energy is taken into account by way of effective
masses and momenta and the Fermi motion of the correct density by
the chemical potential $\mu^*$. Thus this temperature is a local thermodynamic 
temperature, which in the following we denote as $T_{loc}$. 

Expression (\ref{fermi}) is appropriate for a system in local 
equilibrium.  In a heavy ion collision this is generally not the case 
and would lead to an interpretation of collective kinetic energies in terms of 
temperature. To account for anisotropy effects, e.g. in ref. 
\cite{ksg95} longitudinal and 
perpendicular temperatures have been introduced. 
In our approach we model anisotropic momentum distributions 
by counter-streaming 
or 'colliding' nuclear matter \cite{fuchs97,sehn96,tuebingen95}, 
i.e., by a superposition of two Fermi distributions   
$ n^{(12)} = n^{(1)} + n^{(2)} -\delta n^{(12)}$,
where $\delta n^{(12)} = \sqrt{n^{(1)} \cdot n^{(2)} }$ guarantees the 
validity of the Pauli principle 
and provides a smooth transition to one equilibrated system. 
In \cite{fuchs97,gait99} it has been 
demonstrated that this ansatz 
allows a reliable description of the participant and spectator 
matter at each stage of the reaction. 

Experimentally much of the information about the thermodynamical behaviour in 
heavy ion collisions originates from the analysis of fragment observables. Thus 
also in the present 
analysis we will need to generate and analyze 
fragments. The correct and practical procedure how to properly decribe fragment 
production is still very much debated \cite{fluct,guarnera}. Here we 
do not enter into this debate but use the simplest algorithm, namely a 
coalescence model, as we have done and decribed it in ref. 
\cite{gait99}. In brief, we apply phase space coalescence, i.e. nucleons form a 
fragment, if their positions and momenta ($\vec{x}_{i},\vec{p}_{i}$) satisfy
$| \vec{x}_{i}-\vec{X}_f | \leq R_c$ and  
$| \vec{p}_{i}-\vec{P}_f |  \leq P_c$. $R_c,P_c$ are parameters which are 
fitted to reproduce the observed mass distributions and thus guarantee 
a good overall description of the fragment multiplicities. 

Fragment kinetic energy spectra have been analyzed experimentally 
in the Siemens-Rassmussen or blast 
model \cite{eos95,reisdorf,siemens}. In this model the kinetic energies are 
interpreted in terms of a thermalized freeze-out configuration, characterized by 
a common temperature and a radial flow, i.e. by an isotropically expanding 
source. The kinetic energies are given by 
\begin{equation}
\frac{dN}{dE} \sim pE \: \int \beta^2 d\beta n(\beta) \exp(\gamma E/T) 
\times \:  \Big[ \frac{ \sinh{\alpha} }{ \alpha }\left( \gamma+\frac{T}{E} 
\right) - 
\frac{T}{E} \cosh{\alpha} \Big]
\label{fit}
\qquad ,
\end{equation}
where $p$ and $E$ are the center of mass momentum and the total energy of the 
particle with mass $m$, respectively, and  where $\gamma^{-2}=1-\beta^{2}$ 
and $\alpha=\gamma \beta p/T$. 
Various assumptions have been made for the flow profile $n(\beta)$. A good 
parametrization is a Fermi-type function \cite{reisdorf2}. However, the 
results are not very different when using a 
single flow velocity, i.e. $n(\beta) \sim \delta(\beta-\beta_f)$, which we also 
use here for simplicity.  One then has two parameters in the fit, namely 
$\beta_f$ and the temperature parameter in eq.(\ref{fit}), which we call 
$T_{slope}$. It is, of course, not obvious that $T_{slope}$ represents a 
thermodynamical 
temperature.  One of the aims of this investigation is, in fact, to find its 
significance. The expression (\ref{fit}) has been applied to kinetic energy 
spectra of all fragment masses simultaneously, yielding a global 
$T_{slope}(global)$, 
or to each fragment mass separately, giving $T_{slope}(A_f)$. If a global 
description was achieved, it was concluded that a freeze-out configuration 
exists. We will also test this procedure.

In the following we apply the above methods to  central ($b=0$ fm) and 
semi--central ($b=4.5$ fm) $Au + Au$ collisions at $E_{beam}=0.25-0.8$ AGeV. 
This reactions have been 
studied extensively by the ALADIN \cite{poch95,aladin98,aladin99} and 
EOS \cite{eos95} 
collaborations with respect to temperature and phase transitions. In ref. 
\cite{fuchs97} we have previously studied this reaction at one energy, 600 
MeV/A, 
only with respect to local temperatures and thermodynamical instabilities. In 
this work we 
perform the fragment analysis with the blast model to extract and compare slope 
temperatures and we discuss a wider range of 
incident energies \cite{hirsch99}. 

The spectator is that part of the system which has not collided with the other 
nucleus, but which is nevertheless excited due to the shearing--off of part of 
the nucleus and due to absorption of participant particles. In the calculation 
it is identified as those particles which have approximately beam rapidity.
 It was seen in ref. \cite{fuchs97} that it represents a well 
equilibrated piece of nuclear matter at finite temperature.  

In fig. 1 we show 
the evolution with time of the local temperature and the density for the 
spectator at various incident energies. After the time when the spectator is 
fully developed the properties are rather independent of incident energy which 
supports the freeze-out picture. Also after this time the density and 
temperature 
remain rather constant for several tens of fm/c, making it an ideal system in 
order to study the thermodynamical evolution of low-density, finite temperature 
nuclear matter. In ref. \cite{fuchs97} we also determined pressure and studied 
the dependence of pressure on density. We found that after about 45 fm/c the 
effective compressibility $K \sim \partial P/\partial \rho |_T$ became negative 
indicating that the system enters a region of spinodal instability and 
should subsequently break up into fragments. At this time we find 
densities of about $\rho \sim 0.4 - 0.5 \rho_0$ and $T \sim 5 - 6$ MeV, which is 
in good agreement with findings of the ALADIN group based on isotope 
thermometers \cite{aladin98,aladin99}.
Recently the ALADIN group has also determined kinetic energy spectra of 
spectator fragments and has extracted slope temperatures using eq.(\ref{fit}). 
It was found that these are typically 10 to 12 MeV higher than 
those measured with the isotope thermometer. 

Applying the coalescence model to the spectator we obtain kinetic energy spectra 
as shown in fig.2 at 600A MeV for nucleons ($A_f=1$) and for fragments with 
$A_f \ge 2$ separately. We fitted these spectra with the model of eq. 
(\ref{fit}) in the rest frame of the spectator ($\beta_f=0$). 
The $A_f \ge 2$  spectrum is 
well fitted with a temperature of $T_{slope} \sim (17 \pm 2)$ MeV. The nucleon 
spectrum, on the other hand, shows a  two-component structure, as was also 
observed experimentally in ref. \cite{aladin99}. 
It is dominated by a low energy 
component with $T_{slope,low} = (7.3 \pm 3.5)$ MeV. The high energy 
component has rather poor statistics in our calculation and  we interpret 
it as nucleons from the 
participant that have entered the spectator region. The slope temperature 
of the low 
energy component is close to the local temperature  $T_{loc}=(5 - 6)$ MeV as 
discussed above with respect to fig. 1. 
Thus for nucleons both methods of temperature determination consistently 
are seen to yield about 
the same result. In fact, they should not neccessarily be identical, 
since $T_{loc}$ 
is determined fitting the momentum distribution by a Fermi function while 
$T_{slope}$ in eq. (\ref{fit}) is based on a Maxwell-Boltzmann distribution, 
which are not the same at such low temperatures.

On the other hand the slope temperatures of the fragments are considerably 
higher than those of the nucleons. In fig. 3 we show the slope 
temperatures separately for the different fragment masses and also the 
local temperature for comparison. There is a rapid 
increase of  $T_{slope}$ with fragment mass which saturates for $A_f \ge 3$ 
around $T_{slope} \sim 17$ MeV, which was the temperature determined in fig. 2. 
The experimental values from ALADIN \cite{aladin99} also shown in fig. 3.
were obtained by analogous blast model fits to the measured spectra. 
It can be seen that the slope temperatures from the theoretical calculations 
and from the 
data agree extremely well. Also the corresponding kinetic energies which 
range from from 23.7 MeV ($A_f =2$) to 28.1 MeV ($A_f =8$) are in good 
agreement with the ALADIN data. 

At first sight it is surprising that $T_{slope}$  for nucleons and 
fragments differ from each other and also from $T_{loc}$. The difference 
 has been interpreted in ref.\cite{aladin99} in terms 
of the Goldhaber model \cite{goldhaber}, as it has been applied to fragmentation 
by Bauer \cite{bauer}. When a system of fermions of given density and 
temperature suddenly breaks up the fragment momenta are approximately 
given by the sum of the momenta of the nucleons 
before the decay. For heavier fragments the addition of momenta can be 
considered as a stochastic process which via the central limit theorem leads to 
Gaussian energy distributions which resemble Maxwell distributions and thus 
contribute to the slope temperature. 
As discussed by Bauer and also in ref. 
\cite{aladin99} this effect increases the slope temperatures by an amount which 
is of the order of the difference between the isotope and the slope 
temperatures. 

We wanted to see whether a similar effect can 
explain the mass dependence seen in fig. 3. We 
therefore initialized statistically a system of the mass 
and temperature of the spectator, and subjected it to the 
same fragmentation procedure (coalescence) and to the same fit by eq. 
(\ref{fit}) as we did for the heavy ion collision. These slope
temperatures obtained from the 
statistical model are given in fig. 3 as a band, which corresponds to 
initializations between $\rho = 0.3 
\rho_{0}$ and $T = 6$ MeV and  $\rho = 0.4 \rho_{0}$ and $T = 5.5$ MeV,
which cover the range of values in fig. 1. It 
is seen that the model qualitatively explains the increase in the slope 
temperature relative to the local temperature and the increase with fragment 
mass relative to that for nucleons. A similar conclusion was drawn in ref 
\cite{aladin99} using the results from ref. \cite{bauer}. 
This shows that $T_{slope}$  is {\it not} a thermodynamic temperature. 
The difference relative to the thermodynamic temperature can be understood 
from the fact, that to form a fragment the internal kinetic energies of the 
nucleons are limited by the coalescence condition. Since on the average all 
nucleons 
have the same momenta, this means that the collective momentum per nucleon 
of the fragment 
increases relative to the average. This simulates a  higher 
temperature. This effect has been called ''contribution of Fermi motion to 
the temperature''. The purpose of using the Goldhaber model here was to 
demonstrate this 
effect. Whether a Goldhaber model applies to heavy ion collisions, can, of 
course, be debated, but our results are independent of this question.

Thus we seem to understand fairly well the kinetic energy spectra of the 
spectator fragments and we now turn to the participant region. 
The participant zone in a heavy ion collision constitutes another limiting, but 
still simple case for the investigation of the thermodynamical behaviour of 
nuclear matter. In contrast to the spectator zone one expects a 
compression-decompression cycle and thus richer phenomena with respect to 
fragmentation. The situation becomes particularly simple if we look at central 
collisions of symmetric systems which experimentally are selected 
using transverse energy distributions, charged particle multiplicities or 
polar angles near mid-rapidity  \cite{eos95,reisdorf2,fopi}.

We begin to characterize the calculated evolution of a collision for the 
case of $Au + Au$ at 600A MeV. A very well developed radial flow pattern 
appears after 
about 20 fm/c in agreement with findings of other groups 
\cite{aichelin,nemeth97,hombach98}. The pressure in this reaction 
becomes isotropic at about 35 fm/c indicating equilibriation.
The number of collisions drops to small values at about 40 fm/c. This condition 
we shall call (nucleon) freeze-out. Thus equilibration and freeze-out occur 
rather simultaneously. We find a density at this stage of about normal nuclear 
density and a (local) temperature of about $T_{loc} \sim 15$ MeV in the 
mid-plane of the reaction. 

We now also apply the blast model of eq. (\ref{fit}) to fragment spectra 
generated in 
the coalescence model at the end of the collision at about 90 fm/c. The results 
for the slope temperature $T_{slope}$ and the mean velocity $\beta_f$ are 
shown in fig. 4 for a common fit to all fragments with $A_f \ge 2$ for different 
incident energies. These are compared to the corresponding values extracted by 
the EOS \cite{eos95} and FOPI 
\cite{reisdorf,reisdorf2} collaborations by analogous blast model fits to 
charged particle spectra. Our results are in good 
agreement with the temperatures determined by the FOPI collaboration 
\cite{reisdorf2} 
and somewhat lower than those from EOS \cite{eos95}, in 
particular at higher incident energies. For the radial 
flow, the situation tends to be in reverse,  in particular with respect 
to the FOPI results of ref. \cite{reisdorf}. 
This is generally consistent with the findings of other groups. 
E.g. in ref. \cite{daffin96} 
in a similar approach results were obtained for the EOS data, which are 
close to the data for $T_{slope}$ and and above the data for $\beta_{f}$. 
One should keep in mind, however,  
that $T_{slope}$ and $\beta_f$ are not independent fit parameters. Within the 
uncertainities of the description by blast model fits there is qualitative 
agreement between calculation and experiment.

As was done for the spectator we also apply the blast model separately for 
different fragment masses $A_f$. This is shown in fig. 5 at 600 AMeV in the left 
column. We observe that slope temperatures rise and flow velocities fall with 
fragment mass in contrast to the behaviour for the spectator fragments in fig. 3 
where $T_{slope}$ was about constant. 
A similar behaviour has been seen experimentally at 1 A.GeV in ref. \cite{eos95} 
and in calculations in ref. \cite{hombach98}. It can also be deduced from
fragment spectra at 250 A.MeV shown in ref. \cite{reisdorf2}, which yield 
values very close to the ones given here.

The fragment mass dependence of $T_{slope}$ is thus much stronger and qualitatively 
different than that for the spectator seen in fig. 3. 
Thus this behaviour cannot be interpreted as 
fragments originating from a common freeze-out configuration, i.e. from a 
fragmenting source. To arrive at an interpretation we have shown on 
the right column of fig. 5 the local temperatures and flow velocities for 
different times 
before the nucleon freeze-out, i.e. for $t'=t_{freeze-out}-t$, with  
$t_{freeze-out} \sim 35 fm/c$. (We recall that the values at the left of fig. 5 
are obtained at the end of the reaction, i.e. about 90 fm/c.) It is seen that 
for $A_f = 1$ the values at freeze-out are close to the blast model ones, as 
required. 
However, for fragment masses $A_f > 1$ the slope temperatures and 
velocities behave qualitatively very similar to the local temperatures 
and flow velocities at earlier times . 
This would suggest to interpret the fragment temperatures and velocities 
as signifying that 
heavier fragments originate at times earlier than the nucleon freeze-out. 
This may not be unreasonable
since in order to make a heavier fragment one needs higher densities 
which occur at earlier times and hence higher temperatures. 
However, this does not neccessarily imply that the fragments are really 
formed at this time, since fragments could hardly survive such high temperatures, 
as also discussed in ref. \cite{reisdorf2}. But it could mean that these 
fragments carry information about this stage of the collision. In any 
case it means that in the participant region fragments are {\it not} 
formed in a common equlibrated freeze-out configuration, and that in 
such a situation slope temperatures have to be interpreted with great caution.

In summary fragmentation phenomena in heavy ion collisions are studied as a 
means to explore 
the phase diagram of hadronic matter. For this it is neccessary to determine the 
thermodynamical properties of the fragmenting source. One way to do this 
experimentally is to investigate fragment kinetic energy spectra. In theoretical 
simulations the thermodynamical state can be obtained locally in space and time 
from the phase space distribution. In this work we have compared this with the 
information obtained from the generated fragment spectra. We apply this method 
to the spectator and participant regions of relativistic $Au+Au$-collisions. 
We find that the spectator represents a well developed, equilibrated  and 
instable fragmenting source. The difference in temperature determined from 
the local momentum space (or experimentally from the isotope ratios) and from 
the kinetic energy spectra can be attributed to the Fermi motion in the 
fragmenting source as discussed in a Goldhaber model. 
In the participant region the local temperature at the 
nucleon freeze-out and the slope temperature from fragment spectra are 
different from those of the spectator. 
The slope temperatures rise with fragment mass which might indicate 
that the fragments are not formed in a common, equilibrated  source. These 
investigations should be continued using more dynamic methods of fragment 
formation.
\\ \\
We thank the ALADIN collaboration, in particular W. Trautmann and C. Schwarz, 
for helpful discussions. This work was supported in part by the German ministry 
of education and research BMBF under grant no. 06LM868I and grant no. 06TU887.

\newpage
\begin{figure}[h]
\begin{center}
\leavevmode
\epsfxsize = 15cm
\epsffile[70 110 490 460 ]{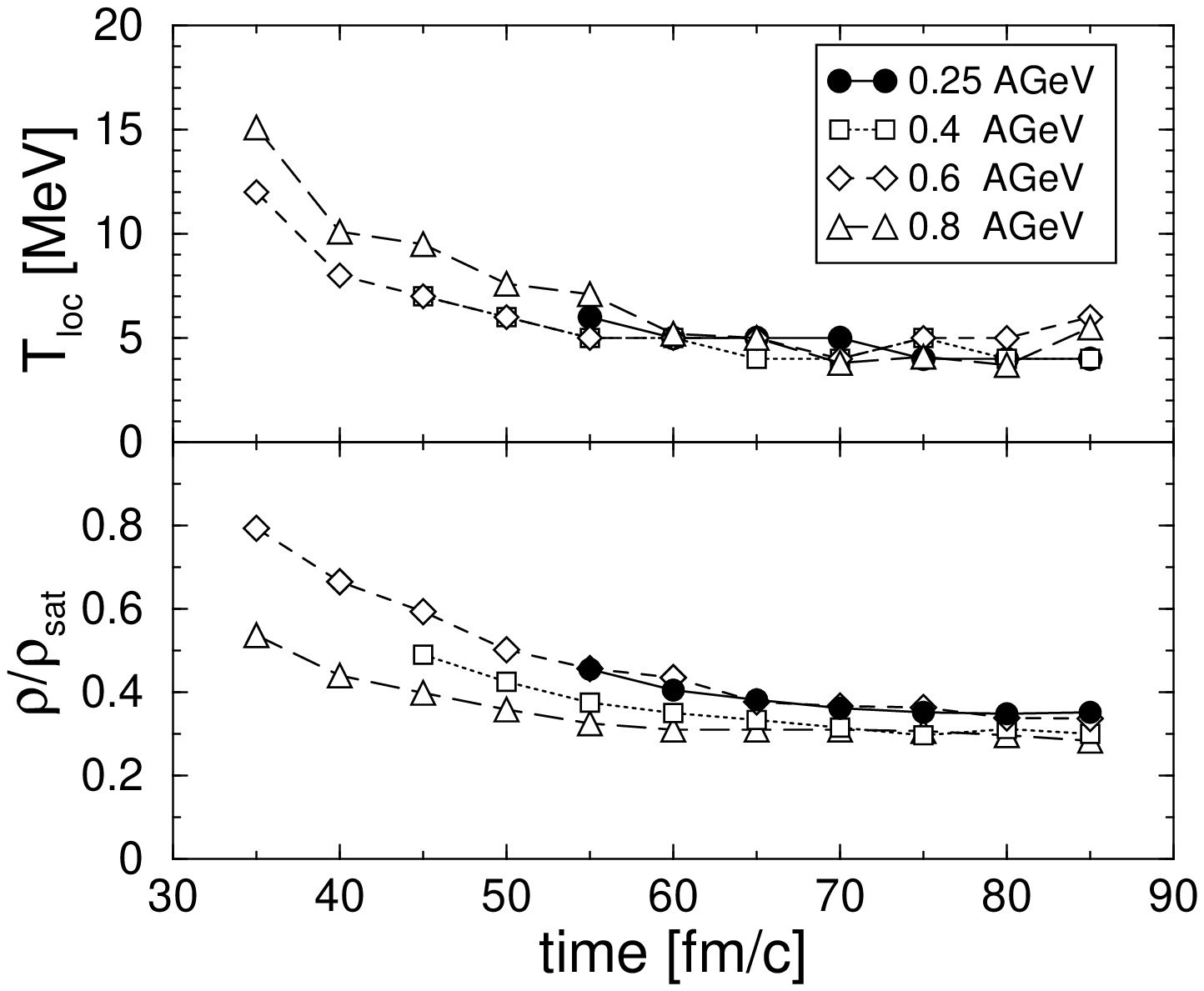}
\end{center}
\caption{ Local temperature (top) and density (bottom) evolution in the 
spectator in semi--central $Au+Au$ reactions (b=4.5 fm) at different beam 
energies as a function of time.}
\label{fig1}
\end{figure}
\clearpage
\begin{figure}[h]
\begin{center}
\leavevmode
\epsfxsize = 15cm
\epsffile[30 30 500 330 ]{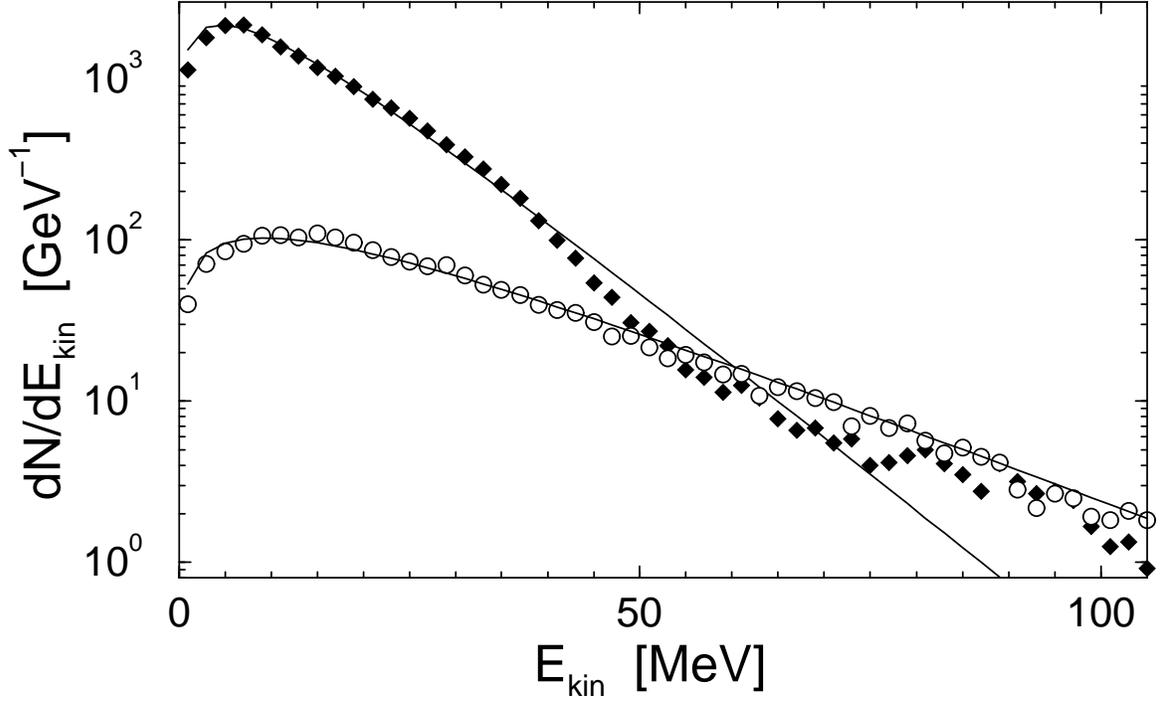}
\end{center}
\caption{ Energy spectra for nucleons ($A_{f}=1$) and fragments 
($A_{f} \geq 2$) (filled diamonds and open circles, respectively) 
from the spectator for $Au+Au$ at 
$E_{beam}=600$ AMeV. Blast model fits are shown (solid curves) which
determine slope temperatures.}
\label{fig2}
\end{figure}
\clearpage
\begin{figure}[b]
\begin{center}
\leavevmode
\epsfxsize = 15cm
\epsffile[20 60 420 290]{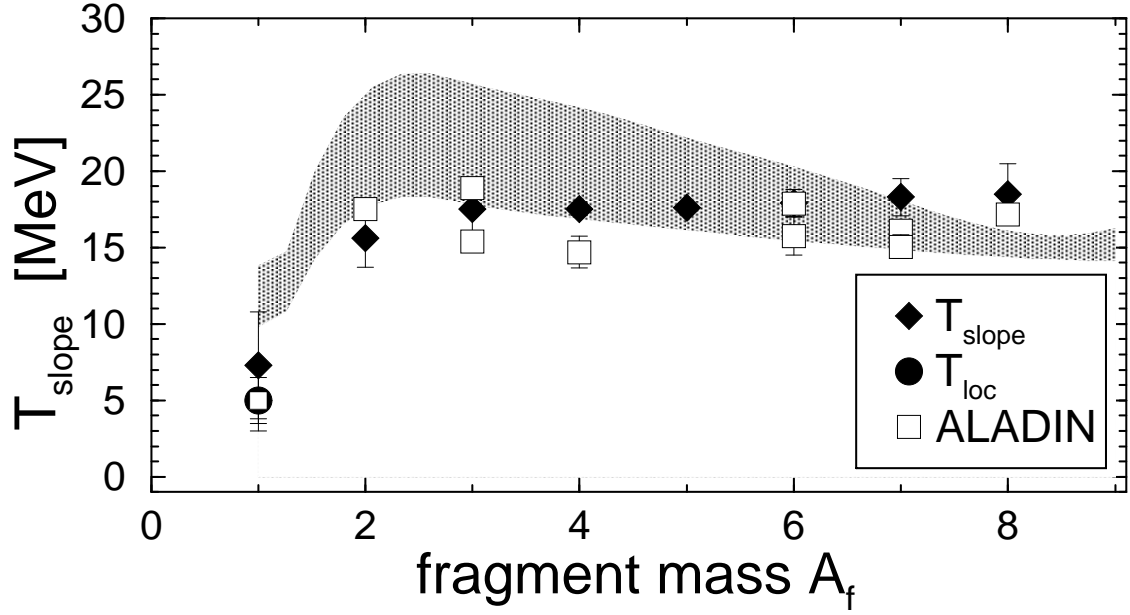}
\end{center}
\caption{ Spectator slope temperatures for different fragment masses $A_{f}$ 
for the reaction as in fig. \protect\ref{fig2} (diamonds). Also shown is the 
nucleon local temperature (circle) and the temperature obtained from a 
statistical model (triangle and gray band, see text).}
\label{fig3}
\end{figure}
\clearpage
\begin{figure}[b]
\begin{center}
\leavevmode
\epsfxsize = 15cm
\epsffile[70 100 360 450]{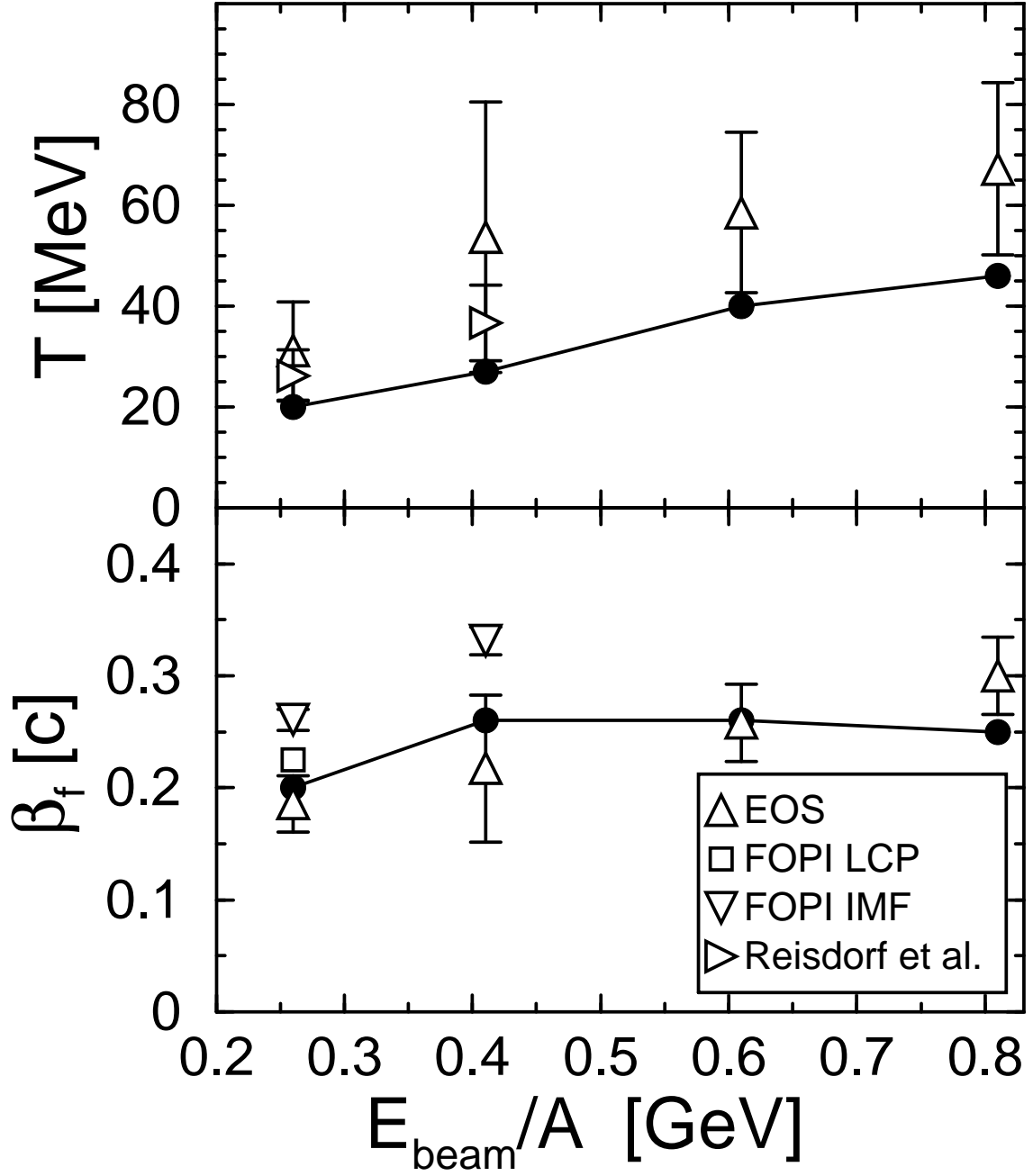}
\end{center}
\caption{Slope temperatures (top) and radial flow velocities (bottom) 
from blast model fits to fragment ($A_f>1$) energy spectra 
for central collisions for different beam energies (filled dots). 
They are compared to data from \protect\cite{eos95,reisdorf,reisdorf2}.}
\label{fig4}
\end{figure}
\clearpage
\begin{figure}[b]
\begin{center}
\leavevmode
\epsfxsize = 15cm
\epsffile[30 30 550 550 ]{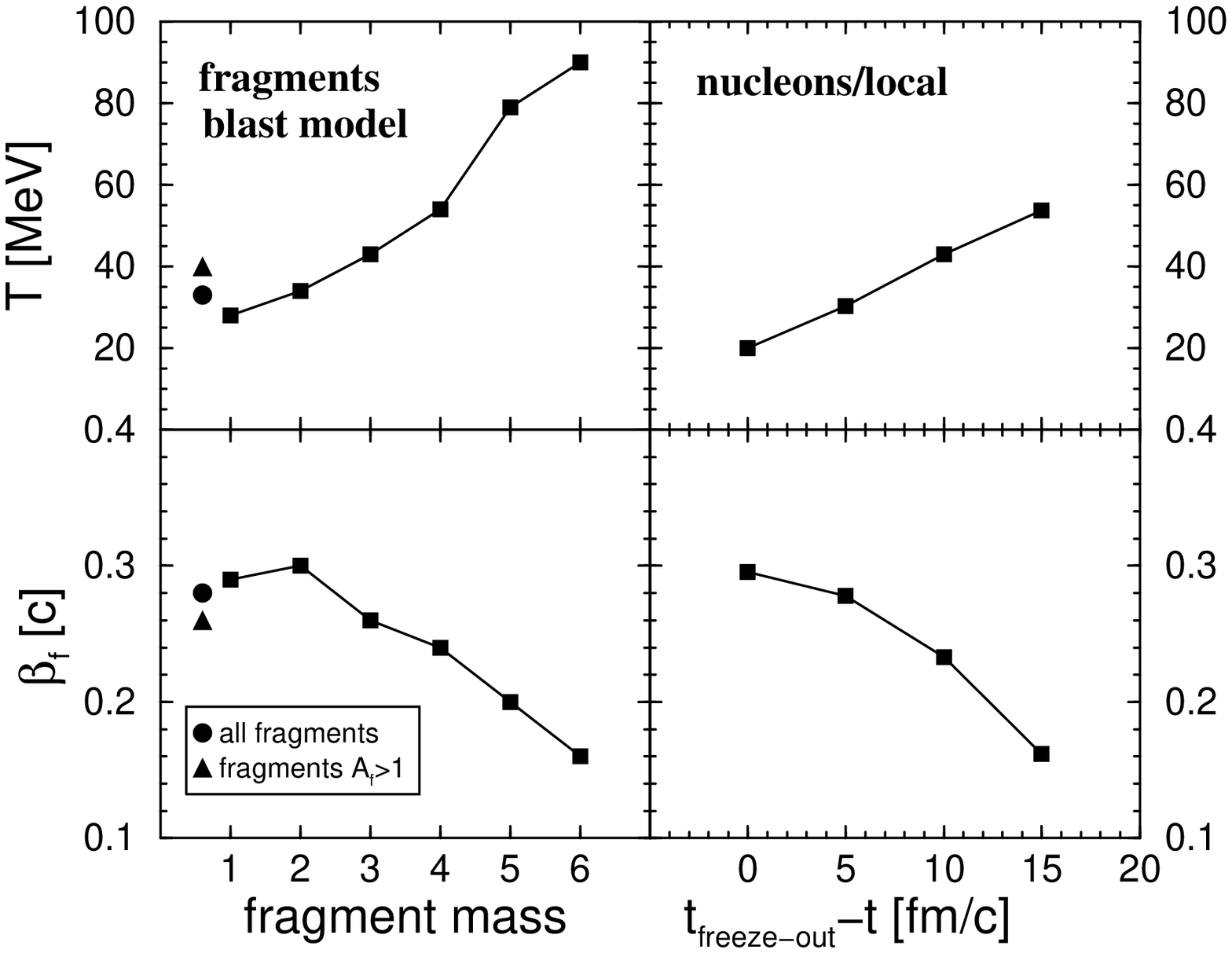}
\end{center}
\caption{
Slope temperatures (upper row) and flow velocities (lower row) for the 
same reaction as in fig. 4 at $E_{beam}=0.6$ AGeV. In the left column 
 for blast model fits for different fragment masses and 
also for $A_f>1$ and for all fragments;
in the right column the local values from the momentum distribution 
at times before the freeze--out (see text).}
\label{fig5}
\end{figure}
\clearpage

\begin{thebibliography}{99}
\bibitem{poch95}
J. Pochodzalla, Prog. Part. Nucl. Phys. {\bf 39} (1997) 443.

\bibitem{aichelin}
J. Aichelin, Phys. Reports {\bf 202} (1991) 233;
R. Nebauer and J. Aichelin, Nucl. Phys. {\bf A 650} (1999) 65;
P. B. Gossiaux, R. Puri, Ch. Hartnack, J. Aichelin, 
Nucl. Phys. {\bf A 619} (1997) 379.
\bibitem{fuchs95}
 C. Fuchs, H.H. Wolter, Nucl. Phys. {\bf A 589} (1995) 732.

\bibitem{nemeth97}
J. Nemeth and G. Papp, Phys. Rev. {\bf C 59} (1999) 1802.

\bibitem{hombach98}
A. Hombach, W. Cassing, S. Teis, U. Mosel, Eur. Phys. J. {\bf A 5} (1999) 157.

\bibitem{daffin96}
F. Daffin, K. Haglin, and W. Bauer, Phys. Rev. {\bf C 54} (1996) 1375.

\bibitem{fuchs97}
C. Fuchs, P. Essler, T. Gaitanos and H.H. Wolter, 
Nucl. Phys. {\bf A626} (1997) 987.

\bibitem{eos95}
M. Lisa and the EOS collaboration, 
Phys. Rev. Lett. {\bf 75} (1995) 2662.

\bibitem{reisdorf}
W.~Reisdorf and H.~G.~Ritter, Annu.~Rev.~Nucl.~Part.~Sci. {\bf 47} (1997) 663, 
and references there.

\bibitem{reisdorf2}
W. Reisdorf and the FOPI Collaboration, Nucl. Phys. {\bf A 612} (1997) 493.

\bibitem{NL2}
B. Bl\"attel, V. Koch, U. Mosel, Rep. Prog. Phys. {\bf 56} (1993) 1.

\bibitem{gait99}
T. Gaitanos, C. Fuchs, and H. H. Wolter, Nucl. Phys. {\bf A650} (1999) 97 .

\bibitem{ksg95}
J. Konopka, H. St\"ocker and W. Greiner, Nucl. Phys. {\bf A 583} (1995) 357. 

\bibitem{sehn96}
L. Sehn, H.H. Wolter, Nucl. Phys. {\bf A 601} (1996) 473;\\
C. Fuchs, L. Sehn, H.H. Wolter, Nucl. Phys. {\bf A 601} (1996) 505.

\bibitem{tuebingen95}
R.K. Puri et al., Nucl. Phys. {\bf A 575} (1995) 733.

\bibitem{fluct}
S. Ajik, et al., Nucl. Phys. {\bf A 513} (1990) 187; J. Randrup et al.,
Nucl. Phys. {\bf A 514} (1990) 339.

\bibitem{guarnera} A. Guarnera, et al., Phys. Lett. {\bf B 373} (1996) 267;
M. Colonna et al., Nucl. Phys. {\bf A 642} (1998) 449.

\bibitem{siemens}
P.~J.~Siemens and J.~O.~Rasmussen, Phys.~Rev.~Lett. {\bf 42} (1979) 880.

\bibitem{aladin98}
V. Serfling and the ALADIN Collaboration, 
Phys. Rev. Lett. {\bf 80} (1998) 3928.

\bibitem{aladin99}
W. Trautmann and the ALADIN Collaboration, 
see \protect\cite{hirsch99}; C. Schwarz, {\it ibid}.

\bibitem{hirsch99}
T. Gaitanos, C. Fuchs, H. H. Wolter, 
Proc. of the International 
Workshop XXVII on Gross Properties of Nuclei and Nuclear 
Excitations, Hirschegg, Austria, 1999.

\bibitem{goldhaber}
A. S. Goldhaber, Phys. Lett. {\bf B 53} (1974) 306; 
Phys. Rev. {\bf C 17} (1978) 2243.

\bibitem{bauer}
W. Bauer, Phys. Rev. {\bf C 51} (1995) 803.

\bibitem{fopi}
F. Rami and the FOPI Collaboration, Nucl. Phys. {\bf A 646} (1999) 367.

\end{thebibliography}
\end{document}